\documentclass{article}
\usepackage{amsfonts}
\usepackage{amsmath}
\usepackage{amssymb}
\usepackage{graphicx}
\newtheorem{theorem}{Theorem}
\newtheorem{acknowledgement}[theorem]{Acknowledgement}

\newtheorem{proposition}[theorem]{Proposition}

\newenvironment{proof}[1][Proof]{\noindent\textbf{#1.} }{\ \rule{0.5em}{0.5em}}

\def\bbbn{{\mathbb N}}

\def\bbbc{{\mathbb C}}

\def\Tr{\mbox{Tr}\,}

\def\tr{\mbox{Tr}\,}

\def\bbbz{{\mathbb Z}}

\def\iH{{\mathcal H}}
\def\iA{{\mathcal A}}

\def\<{\langle}
\def\>{\rangle}

\newcommand{\sm}[1]{^{\hbox{\tiny$\begin{array}{@{}c}{#1}\\[-0.5em]
\smile\end{array}\! \! \! \! \! $}}}
\begin{document}

\title { Notes on the equality in SSA of entropy on CAR algebra}
\author{J. Pitrik\footnote{e-mail: pitrik@math.bme.hu} , V.P. Belavkin\footnote{e-mail: Viacheslav.Belavkin@maths.nottingham.ac.uk}\\School of Mathematical Sciences\\University of Nottingham, NG7 2RD, UK}
\date{}
\maketitle

\begin{abstract}
We prove a necessary and sufficient condition for the states which satisfy strong subadditivity of von Neumann entropy with equality on CAR algebra and we show an example when the equality holds but the state is not separable.
\end{abstract}

\section{Introduction}

A remarkable property of the von Neumann entropy 
\begin{equation}
S(D)=-\Tr D\log D
\end{equation}
of a density operator $D$ ($D>0$ , $\Tr D=1$) on a Hilbert space $\iH$ is the strong subadditivity (SSA) which was proved by Lieb and Ruskai \cite{LR}
\begin{equation}
S(D_{12})+S(D_{23})\ge S(D_{123})+S(D_2).
\end{equation}
with a tripartite state $D_{123}$ on the system $\iH_1\otimes\iH_2\otimes\iH_3$. It is interesting to find the states which saturate the SSA inequality because they are exactly the markovian states for tensor product systems. In \cite{HA} an explicit characterisation of such quantum states was given.

For CAR algebras (we summarize their properties in the next section) the SSA inequality of entropy was showed by Araki and Moriya in \cite{AM} , but the question of the case of the equality is still open. We give an equivalent  condition for the states which is already well-known for tensor product. We show a separable and a non-separable class of states which saturate the SSA inequality.

\section{CAR}

In this section we summarize known properties of the algebra
of the canonical anticommutation relation. The paper of Araki and Moriya \cite{AM} contains all what we need.

Assume that the unital C*-algebra $\iA_I$ is generated 
by the elements $\{a_i: i \in I:=\{1,2,\dots,n\}\}$ which 
satisfy the relations
\begin{eqnarray*}
a_i a_j +a_ja_i & = & 0 \\
a_i a_j^* +a_j^* a_i & = & \delta_{i,j}
\end{eqnarray*}
for $i,j \in I$. It is easy to see that $\iA_{I}$ is the linear
span of the identity and monomials of the form
\begin{equation}\label{E:terms}
A_{i(1)}A_{i(2)}\dots A_{i(k)}, 
\end{equation}
where $1 \le i(1)< i(2) <\dots < i(k) \le n$ and each factor 
$A_{i(j)}$ is one of the four operators $a_{i(j)},a_{i(j)}^*,
a_{i(j)}a_{i(j)}^*, a_{i(j)}^*a_{i(j)}$.  

It is known that $\iA_{I}$ is isomorphic to a matrix algebra 
$M_{2^n}(\bbbc)\simeq M_2 (\bbbc)^{\sm{1}} \otimes \cdots \otimes 
M_2 (\bbbc )^{\sm{n}}$.
Namely, the relations
$$
\begin{array}{lcl@{\qquad}lcl}
e_{11}^{(i)} \, : & =& a_i a_i^*& e_{12}^{(i)} \, : & = & V_{i-1} a_i\\[0.3em]
e_{21}^{(i)} \, : & =& V_{i-1} a_i^*& e_{22}^{(i)} \, : & = & a_i^*
a_i\\[0.1em]
\end{array}
$$

$$
V_i := \prod_{j=1}^i (I-2 a_j^* a_j )
$$
determine a family of mutually commuting $2 \times 2$ matrix units for $i 
\in I$. Since
$$ 
a_i = \prod_{j=1}^{i-1} \left( e_{11}^{(j)} -e_{22}^{(j)} \right)
e_{12}^{(i)} ,
$$
the above matrix units generate $\iA_{I}$ and give an isomorphism between
$\iA_{I}$ and $M_2 (\bbbc ) \otimes \cdots \otimes M_2 (\bbbc )$:
\begin{equation}\label{E:izo}
 e_{i_1 j_2}^{(1)} e_{i_2 j_2}^{(2)} \ldots e_{i_n j_n }^{(n)}
\longleftrightarrow e_{i_1 j_1} \otimes e_{i_2 j_2} \otimes \cdots \otimes
e_{i_n j_n} .
\end{equation}
(Here $e_{ij}$ stand for the standard matrix units in $M_2 (\bbbc )$.) It
follows from this isomorphism that $\iA_I$ has a unique tracial state
$\tau$.

Let $J \subset I$. There exists a unique automorphism $\Theta_J$ of 
$\iA_I$ such that
\begin{eqnarray*}
&& \Theta_J (a_i) = -a_i \mbox{\ and\ } 
\Theta_J (a_i^*)=- a_i^* \quad (i \in I) \\
&&\Theta_J (a_i) = a_i \mbox{\ and\ } 
\Theta_J (a_i^*)=a_i^* \quad (i \notin I).
\end{eqnarray*}
In particluar, we write $\Theta$ instead of $\Theta_I$. The {\it odd}
and {\it even parts} of $\iA_I$ are defined as
\begin{equation}
\iA_I^+:=\{ a \in \iA_I: \Theta(a)=a\},\quad 
\iA_I^-:=\{ a \in \iA_I: \Theta(a)=-a\}.
\end{equation}
$\iA_I^+$ is a subalgebra but $\iA_I^-$ is not.
The graded commutation relation for CAR algebras is well known:
if  $A\in\iA (K)$ and $B\in\iA (L)$ where $K\cap L=\emptyset$ , then $AB=\epsilon (A,B)BA$ where
\begin{equation}
\epsilon(A,B)=\left\{
\begin{array}{cc}
 {-1}&\mbox{if A and B are odd}\\
 {+1}&\mbox{otherwise.}
\end{array}
\right.
\end{equation}
The map
\begin{equation}
A \mapsto \frac{1}{2}\big(A+\Theta(A)\big)
\end{equation}
is a conditional expectation of $\iA_I$ onto $\iA_I^+$. We have
$\iA_I=\iA_I^+ + \iA_I^-$ and $\Theta$ leaves the trace $\tau$
invariant.

Let $J \subset I$. Then $\iA_J \subset \iA_I$ and there exists a
unique conditional expectation $E^I_J:\iA_I \to \iA_J$ which
preserves the trace. This fact follows from generalities about
conditional expectations or the isomorphism (\ref{E:izo}). 
Inspite of these, it is useful to have a construction for 
$E^I_J$. The C*-algebra generated by the commuting subalgebras
$\iA_J$ and  $\iA_{I\setminus J}^+$ is isomorphic to their 
tensor product. We have a conditional expectation
\begin{equation}
F_1:\iA_I \to \iA_J \otimes\iA_{I\setminus J}^+, 
\quad F_1(A)=\frac{1}{2}\big(A+\Theta_{I\setminus J}(A)\big)
\end{equation}
and another
\begin{equation}
F_2:\iA_J \otimes\iA_{I\setminus J}^+ \to \iA_J , 
\quad F_2(A\otimes B)=\tau(B)\,A.
\end{equation}

The composition $F_2\circ F_1$ is $E^I_J$. To have an example, assume
that $I=[1,4]$, $J=[1,2]$ and consider the action of the above conditional
expectations on terms like (\ref{E:terms}). $F_1$ keeps 
$a_1a_2^*a_2a_3^*a_4$ fixed and $F_2$ sends it to $a_1a_2^*a_2\tau(a_3^*)
\tau(a_4)=0$. Moreover, $E^I_J$ sends $a_1a_2^*a_2a_3 a_3^*a_4^*a_4$
to $a_1a_2^*a_2\tau(a_3 a_3^*)\tau(a_4^*a_4)$. We make here two remarks.
First, we have benefitted from the product property of the trace:
If $A\in \iA_{J_1}$, $B\in \iA_{J_2}$ for disjoint subsets $J_1$ and $J_2$
of $I$, then
\begin{equation}
\tau(AB)=\tau(A)\tau(B)\, .
\end{equation}
Moreover, for arbitrary subsets $J_1,J_2 \subset I$
\begin{equation}
E^I_{J_1}|\iA_{J_2}=E^{J_2}_{J_1\cap J_2}
\end{equation}
holds. This means that we have a commuting square:$ E^{I}_{J_1}$
\begin{center}
\begin{picture}(60,30)
\put(30,1){\makebox(0,0)[cc]{$\iA_{J_1 \cap J_2}$}}
\put(30,29){\makebox(0,0)[cc]{$\iA_{I}$}}
\put(2,15){\makebox(0,0)[cc]{$\iA_{J_1}$}}
\put(58,15){\makebox(0,0)[cc]{$\iA_{J_2}$}}
\put(25,4){\vector(-2,1){17}}
\put(24,26){\vector(-2,-1){16}}{$E^{I}_{J_1}$}
\put(52,12.5){\vector(-2,-1){17}}
\put(52,18){\vector(-2,1){16}}
\put(45,6.5){\makebox(0,0)[lc]{$E^{J_1}_{J_1 \cap J_2}$}}
\put(44,25.5){\makebox(0,0)[lc]{$\,$}}
\put(14,6.5){\makebox(0,0)[rc]{$\,$}}
\put(16,25.5){\makebox(0,0)[rc]{$E^{I}_{J_1}$}}
\end{picture}
\end{center}

\section{Equality in SSA}

Araki and Moriya proved that the strong subadditive property of the von Neumann entropy also holds for CAR systems \cite{AM} ie.
\begin{theorem}
For finite subsets $I$ and $J$ of ${\bbbz}^{\nu}$, the strong subadditivity (SSA) of $S$ holds for any state $\psi$ of $\iA$:
$$S({\psi}_{I\cup J})-S({\psi}_{I})-S({\psi}_{J})+S({\psi}_{I\cap J})\le 0$$
where ${\psi}_{K}$ denotes the restriction of $\psi$ to $\iA (K)$, and $S$ is the von Neumann entropy.
\end{theorem}
To investigate the condition of the equality we need the following theorem, which can be found as Theorem 3.8 in the monograph of Ohya and Petz \cite{OP}.

\begin {theorem}
Let $\iA$ be a finite quantum system, and $A,B,C\in\iA^{sa}$.Then the following inequality holds:
\begin{equation}
\tr{e^C T_{\exp (-A)}(e^B)}\ge \tr{e^{A+B+C}}
\end{equation}
where $T_A (K)=\int_0^{\infty} (t+A)^{-1}K(t+A)^{-1}\,\mathrm{d}t$.
\end{theorem}
Now we can prove the following
\begin{proposition}
Equality holds in SSA iff
\begin{equation}
\log D +\log D_{I\cap J}=\log D_{I}+\log D_{J}
\end{equation}
where $D$ is the density matrix of $\omega$ and $D_K$ is its restriction by the conditional expectation onto the subalgebra $\iA (K)$. 
\end{proposition}

\begin{proof}
From the proof of the SSA we can observe that the equality holds if and only if
\begin{equation}
S(\omega, \omega\circ E^{I\cup J}_{I}) = S(\omega\circ E^{I\cup J}_{J}, \omega
\circ E^{I\cup J}_{I\cap J})\
\end{equation}
where $\omega$ is a faithful normal state on $\iA(I\cup J)$. We show that this condition is equivalent to the above mentioned one.

For the sufficiency let us consider a $D$ density matrix on $\iA(I\cup J)$, ie. $D\in\iA(I\cup J)$, $D>0$ and $\tau (D)=1$.

\begin{eqnarray*}
S(\omega, \omega\circ E^{I\cup J}_{I}) - S(\omega\circ E^{I\cup J}_{J}, \omega
\circ E^{I\cup J}_{I\cap J})&=&\\
\omega (\log D-\log E^{I\cup J}_{I}(D))-\omega\circ E^{I\cup J}_{J}(\log E^{I\cup J}_{J}(D)-\log  E^{I\cup J}_{I\cap J}(D))&=&\\
\omega (\log D-\log E^{I\cup J}_{I}(D))-\omega(\log E^{I\cup J}_{J}(D)-\log  E^{I\cup J}_{I\cap J}(D))&=&\\
\omega (\log D-\log D_{I}-\log D_{J}+\log D_{I\cap J})&=&0
\end{eqnarray*}
where we used that $ E^{I\cup J}_{J}(D)$ and $ E^{I\cup J}_{I\cap J}(D)$ are  elements of $\iA_{J}$. This gives us the sufficiency.
For the necessity we have

\begin{eqnarray*}
-S(D)+S(D_{I})-S(D_{I\cap J})+S(D_{J})&=&\\
\tau(D(\log D-(\log D_{I}-\log D_{I\cap J}+\log D_{J}))&\ge&\\
\tau(D-\exp (\log D_{I}-\log D_{I\cap J}+\log D_{J}))
\end{eqnarray*}
where we used the Klein inequality $\tau\left( A(\log A-\log B)\right)\ge \tau (A-B)$ and equality holds if and only if $A=B$.

By using the theorem above with $A=-\log D_{I\cap J}$ ,$B=\log D_{J}$ , and $C=\log D_{I}$ we get the following inequality:

\begin{equation}
\tau(\exp (\log D_{I}-\log D_{I\cap J}+\log D_{J}))\le\tau\left(\int_0^{\infty}D_{I} (tI+D_{I\cap J})^{-1} D_{J}(tI+D_{I\cap J})^{-1}\,\mathrm{d}t\right)
\end{equation}

By using that the conditional expectation is trace preserving ie. $\tau =\tau \circ E$ and the property $E^{M}_{N}(ABC)=AE^{M}_{N}(B)C$ where $A,C\in N$ and $B\in M$ we get

\begin{eqnarray*}
\tau \left(\int_0^{\infty}D_{I} (tI+D_{I\cap J})^{-1} D_{J}(tI+D_{I\cap J})^{-1}\,\mathrm{d}t\right)&=&\\
\tau \circ E^{I\cup J}_{I} \left(\int_0^{\infty}D_{I} (tI+D_{I\cap J})^{-1} D_{J}(tI+D_{I\cap J})^{-1}\,\mathrm{d}t\right)&=&\\
\tau \left(\int_0^{\infty}D_{I} (tI+D_{I\cap J})^{-1}E^{I\cup J}_{I}\left( D_{J}\right)(tI+D_{I\cap J})^{-1}\,\mathrm{d}t\right)&=&\\
\tau \left(\int_0^{\infty}D_{I} (tI+D_{I\cap J})^{-1} D_{I\cap J}(tI+D_{I\cap J})^{-1}\,\mathrm{d}t\right)&=&\\
\tau \circ E^{I\cup J}_{J} \left(\int_0^{\infty}D_{I} (tI+D_{I\cap J})^{-1} D_{I\cap J}(tI+D_{I\cap J})^{-1}\,\mathrm{d}t\right)&=&\\
\tau \left(\int_0^{\infty}E^{I\cup J}_{J}\left(D_{I}\right) (tI+D_{I\cap J})^{-1} D_{I\cap J}(tI+D_{I\cap J})^{-1}\,\mathrm{d}t\right)&=&\\
\tau \left(\int_0^{\infty}D_{I\cap J} (tI+D_{I\cap J})^{-1} D_{I\cap J}(tI+D_{I\cap J})^{-1}\,\mathrm{d}t\right)&=&\\
\tau \left(\int_0^{\infty}\left(D_{I\cap J}\right)^2 (tI+D_{I\cap J})^{-2}\,\mathrm{d}t\right)&=&\tau\left( D_{I\cap J}\right)
\end{eqnarray*}
In the last step we used the fact
\begin{equation}
\int_0^{\infty}\frac{{\lambda}^2}{(t+\lambda )^2}\,\mathrm{d}t=\lambda
\end{equation}
Substituted this result into the inequality above, we have

\begin{equation}
\tau(D-\exp (log D_{I}-\log D_{I\cap J}+\log D_{J}))\ge \tau (D)-\tau (D_{I\cap J})=0
\end{equation}

Using the necessary and sufficent condition for the equality in the Klein inequality, we get

\begin{equation}
D=\exp(log D_{I}-\log D_{I\cap J}+\log D_{J})
\end{equation}

but $D=\exp (\log D)$ which give us the necessity and our proof is complete.
\end{proof}

Araki and Moriya proved the following theorem in \cite{AM2}.
\begin{theorem}
Let $I_1$,$I_2$,$\dots$ be an arbitrary number of mutually disjoint subsets of $\bbbn$ and $\omega_i$ be given state of $\iA(I_i)$ for each $i$. A product state extension of $\omega_i$ $i=1,2,\dots$ exists if and only if all states $\omega_i$ except at most one are even. It is unique if it exists.
\end{theorem}
With the help of this theorem we can prove easily the folloving
\begin{proposition}
If $\omega$ is a separable state for $\iA(I\setminus J)$ , $\iA(I\cap J)$ , $\iA(J\setminus I)$ ie.
\begin{equation}
\omega=\sum_i\lambda_i\omega_{1,i}\circ\omega_{2,i}\circ\omega_{3,i}
\end{equation}  
with $\lambda_i>0$, $\sum_i\lambda_i=1$ where  $\omega_{1,i}$ , $\omega_{2,i}$ and $\omega_{3,i}$ are states on $\iA(I\setminus J)$ , $\iA(I\cap J)$ and $\iA(J\setminus I)$ respectively with orthogonal supports, then $\omega$ satisfy the SSA of entropy with equality.
\end{proposition}
\begin{proof}
If the product state extension  $\omega_{1,i}\circ\omega_{2,i}\circ\omega_{3,i}$ exists, then among the marginal states $\omega_{1,i}$ , $\omega_{2,i}$ , $\omega_{3,i}$ at least two must be even for all $i$. It means the same condition for their density matrices $D_{1,i}$ , $D_{2,i}$ , $D_{3,i}$ so they commute with each other by the graded commutation relation. The density matrix of $\omega$ is given by
\begin{equation}
D=\sum_i\lambda_iD_{1,i}D_{2,i}D_{3,i}
\end{equation}
and its restrictions to $I$ , $J$ and $I\cap J$ are $D=\sum_i\lambda_iD_{1,i}D_{2,i}$ , $D=\sum_i\lambda_iD_{2,i}D_{3,i}$ and $D=\sum_i\lambda_iD_{2,i}$ , respectively. For their entropies we have
\begin{eqnarray*}
S(D)=S(\sum_i\lambda_iD_{1,i}D_{2,i}D_{3,i})&=&H(\lambda_i)+\sum_i\lambda_iS(D_{1,i}D_{2,i}D_{3,i})=\\
&&H(\lambda_i)+\sum_i\lambda_i\left(S(D_{1,i})+S(D_{2,i})+S(D_{3,i})\right)
\end{eqnarray*} 
where $H(\lambda_i)$ is the Shannon entropy of the classical distribution $\lambda_i>0$ , $\sum_i\lambda_i=1$ and in the last step we have used the fact that the marginal densities commute with each other. Similarly we have
\begin{eqnarray}
S(D_I)&=&H(\lambda_i)+\sum_i\lambda_i\left(S(D_{1,i})+S(D_{2,i})\right)\\
S(D_J)&=&H(\lambda_i)+\sum_i\lambda_i\left(S(D_{2,i})+S(D_{3,i})\right)\\
S(D_{I\cap J})&=&H(\lambda_i)+\sum_i\lambda_iS(D_{2,i})
\end{eqnarray}
Substituting the entropies, the equality in SSA is hold.
\end{proof}

Contrary to tensor product systems in CAR systems there are nonseparable states which satisfy the SSA of entropy with equality as the next result shows.

\begin{proposition}
Let consider the set partition $I\cap J={\cup}_i K_i$ where $K_i\cap K_j=\emptyset$ if $i\ne j$. The following density matrix satisfies the SSA with equality
\begin{equation}
D={\sum}_i{\alpha}_iA_iB_iC^+
\end{equation}
where $A_i\in\iA(I\setminus J)$ , $B_i\in\iA(K_i)$ and $C^+\in{\iA (J\setminus I)}^{+}$  (independent on $i$) are monomials and ${\alpha}_i$ are normalization constants. If $B_i$ is odd and $A_i$ is even, $A_i$ must be a product of two disjoint odd elements, i.e. there exist $L^1_i$, $L^2_i$ disjoint sets with $L^1_i\cup L^2_i=I\setminus J$ and there exist $a^1_i\in\iA(L^1_i)^{-}$ and $a^2_i\in\iA(L^2_i)^{-}$ such that $A_i=a^1_ia^2_i$. 
\end{proposition}

\begin{proof}
We can observe that the density matrix above can contain the product of odd monomials living in disjoint subsets, so the necessary condition of the separability does not hold.

From the form of $D$ we have the following reduced density matrices
\begin{eqnarray}
D_{I}&=&E^{I\cup J}_I={\sum}_i{\alpha}_iA_iB_i\tau (C^+)\\
D_{J}&=&E^{I\cup J}_J={\sum}_i{\alpha}_i\tau (A_i)B_i(C^+)\\
D_{I\cap J}&=&E^{I\cup J}_{I\cap J}={\sum}_i{\alpha}_i\tau(A_i)B_i\tau (C^+)
\end{eqnarray}
We recall the graded commutation relation:
if  $A\in\iA (K)$ and $B\in\iA (L)$ where $K\cap L=\emptyset$ , then $AB=\epsilon (A,B)BA$ where
\begin{equation*}
\epsilon(A,B)=\left\{
\begin{array}{cc}
 {-1}&\mbox{if A and B are odd}\\
 {+1}&\mbox{otherwise}
\end{array}
\right.
\end{equation*}
We show that $D_I$ and $D_J$ commute.
\begin{eqnarray*}
D_ID_J&=&\left({\sum}_i{\alpha}_iA_iB_i\tau (C^+)\right)\left({\sum}_j{\alpha}_j\tau (A_j)B_jC^+\right)\\
&=&{\sum}_{i,j}{\alpha}_i{\alpha}_j\tau (C^+)\tau (A_j)\epsilon(B_i,B_j)\epsilon(A_i,B_j)\epsilon(B_i,C^+)\epsilon(A_i,C^+)B_jC^+A_iB_i\\
&=&\left({\sum}_j{\alpha}_j\tau (A_j)B_jC^+\right)\left({\sum}_i{\alpha}_iA_iB_i\tau (C^+)\right)=D_JD_I
\end{eqnarray*}
where we used that $\epsilon(B_i,C^+)=\epsilon(A_i,C^+)=1$ since $C^+$ is even. If $B_j$ is even we have $\epsilon(A_i,B_j)=\epsilon(B_i,B_j)=1$. If $B_j$ is odd and $A_j$ is even, our construction gives $A_j=a^1_ja^2_j$, where $a^1_j$ and $a^2_j$ are odd elements living in disjoint algebras and by using the fact that $\tau$ is an even product state we have $\tau(A_j)=\tau(a^1_j)\tau(a^2_j)=0$. If $A_j$ is odd we have $\tau(A_j)=0$ immediately.

A similar computation shows that $D$ and $D_{I\cap J}$ also commute.
\begin{eqnarray*}
DD_{I\cap J}&=&\left({\sum}_i{\alpha}_iA_iB_iC^+\right)\left({\sum}_j{\alpha}_j\tau(A_j)B_j\tau (C^+)\right)\\
&=&{\sum}_{i,j}{\alpha}_i{\alpha}_j\tau (C^+)\tau (A_j)\epsilon(C^+,B_j)\epsilon(B_i,B_j)\epsilon(A_i,B_j)B_jA_iB_iC^+\\
&=&\left({\sum}_j{\alpha}_j\tau(A_j)B_j\tau (C^+)\right)\left({\sum}_i{\alpha}_iA_iB_iC^+\right)=D_{I\cap J}D
\end{eqnarray*}
Since $\epsilon(C^+,B_j)=1$ because $C^+$ is even. If $B_j$ is even $\epsilon(A_i,B_j)=\epsilon(B_i,B_j)=1$. If $B_j$ is odd and $A_j$ is even we have $\tau(A_j)=\tau(a^1_j)\tau(a^2_j)=0$ by our construction. If $A_j$ is odd $\tau(A_j)=0$ is hold immediately.

Since $C^+$ and $B_j$ commute for all $j$ it is easy to see that
\begin{equation}
DD_{I\cap J}=D_ID_J
\end{equation}
wich gives
\begin{equation}
\log\left( DD_{I\cap J}\right)=\log \left(D_ID_J\right).
\end{equation}
By using our commutation relations proved above, we have
\begin{equation}
\log D+\log D_{I\cap J}=\log D_I+\log D_J
\end{equation}
which is equivalent condition to the equality in the SSA.
\end{proof}

With this we gave an example for non-separable states which satisfy the SSA inequality of entropy with equality.

\begin{acknowledgement}
The first author would like to thank D\'enes Petz, Mil\`an Mosonyi and Bal\'azs D\'ora for helpful discussions and acknowledges the support of EU Research Training Network on Quantum Probability and Applications.
\end{acknowledgement}

\end{document}